\documentclass[11pt,a4paper]{article}
\usepackage{moriond}
\usepackage{graphicx}
\usepackage{cite}
\usepackage{xspace}
\usepackage[latin1]{inputenc}

\def\be{\begin{equation}}
\def\ee{\end{equation}}
\def\bea{\begin{eqnarray}}
\def\eea{\end{eqnarray}}

\newcommand{\sherpa}{S\scalebox{0.8}{HERPA}\xspace}
\newcommand{\amegic}{A\scalebox{0.8}{MEGIC++}\xspace}

\begin{document}
\vspace*{4cm}
\title{New trends in modern event generators}

\author{Tanju Gleisberg$^{(1)}$, Stefan H\"oche$^{(2)}$,
        Frank Krauss$^{(2)}$, Radoslaw Matyszkiewicz$^{(1)}$,
        Marek Sch\"onherr$^{(1)}$, Steffen Schumann$^{(1)}$, 
        Frank Siegert$^{(1)}$, Jan Winter$^{(1)}$}
\address{$^{(1)}$Institut f\"ur Theoretische Physik, TU Dresden
         {\bf D-01062} Dresden, Germany\\
         $^{(2)}$Institute for Particle Physics Phenomenology,
         Department of Physics, University of Durham,
         Durham DH1 3LE, United Kingdom}

\maketitle
\abstract{Some features of modern simulation tools for high-energy physics
           are reviewed.} 

\section{Introduction: The next generation of event generators}

\noindent
In the past decades, event generators have become increasingly important
for the planning of collider experiments and analyses of their data.  In the
LHC era this trend will become even more pronounced, since many of the
interesting signals expected at the LHC - such as signals for the Higgs
boson or alternative electroweak symmetry breaking mechanisms, supersymmetry, 
etc.\ - are severely hampered by large backgrounds, with a significant 
influence of QCD.  Thus, the success of the LHC probably rests
on a precise understanding of these backgrounds.  Examples for this include
the effect of the central jet-veto in vector boson fusion, producing the 
Higgs boson, and in multi-jet backgrounds to SUSY searches.  In view of this,
it is obvious that many of the old tools need to be replaced by newer, and
better ones, such as Pythia8 \cite{pythia8}, Herwig++ \cite{Gieseke:2003hm}, and 
\sherpa \cite{Gleisberg:2003xi}.  Their ongoing construction in fact reflects 
increased experimental needs.  In many cases, they therefore incorporate new, better 
simulation methods, many of which are connected to the systematic inclusion 
of higher-order QCD-corrections.

\section{Parton level: Calculation of signal and background}

\begin{table}
  \begin{center}
  {\footnotesize
  \begin{tabular}{|l||r|r|r|}
  \hline
  Process       & Cross section             & time (helicity) & time (MHV)     \\\hline
  $jj\to jj$    & 745.85 $\mu$b$\pm 0.10\%$ & 66 s            & 44 s           \\\hline 
  $jj\to jjj$   & 81.274 $\mu$b$\pm 0.20\%$ & 1400 s          & 166 s          \\\hline 
  $gg\to gggg$  & 10.145 $\mu$b$\pm 0.23\%$ & 90 ks           & 0.6 ks         \\\hline 
  $jj\to jjjj$  & 23.208 $\mu$b$\pm 0.26\%$ & 210 ks          & 5.8 ks         \\\hline 
  $gg\to ggggg$ & 2.6915 $\mu$b$\pm 0.15\%$ & -               & 17  ks         \\\hline 
  $jj\to jjjjj$ & 7.3294 $\mu$b$\pm 0.17\%$ & -               & 122 ks         \\\hline
  \end{tabular}
  \caption{Performance of different calculational methods for multi-leg QCD matrix elements.  
           Time is given for the calculation of 310000 phase space points.  Clearly, the 
           CSW recursion relations (here labeled with MHV) are superior in performance;  the fact 
           that only up to two quark lines have been included in the respective algorithm,
           has negligible influence on the final result.}
  \label{Tab:1}
  }
  \end{center}
\end{table}     
\noindent
Many of the apparent improvements of current event generators rely on the inclusion of 
higher-order corrections; one method is to use multi-leg tree-level matrix elements (MEs) 
as a base for simulation.  There are a number of such tools on the market, which either 
evaluate Feynman graphs using the helicity method (for instance \cite{Maltoni:2002qb}) 
or recursion relations, e.g.\ \cite{Mangano:2002ea}.  However, apart from being able to 
calculate the MEs quickly, also the ability to integrate efficiently over the final state 
particles' phase space is a major obstacle for a satisfying performance of such tools.  

\noindent
In the following, some results addressing both issues, are presented, serving as 
illustrative examples for current performances.  In Tab.\ \ref{Tab:1}, the performance of 
the helicity method is compared to the performance of the CSW recursion relations 
\cite{Cachazo:2004kj}.  For pure QCD processes both approaches have been implemented in 
the parton level generator \amegic \cite{Krauss:2001iv}, which is a central part of \sherpa.  
\begin{table}
  \begin{center}
  {\footnotesize
  \begin{tabular}{|r|r||r|r||r|r|}      
  \hline
  $gg\to ggg$ & $jj\to jjj$ & $gg\to gggg$ & $jj\to jjjj$ & $gg\to ggggg$ & $jj\to jjjjj$ \\\hline
  5.8$\%$  & 1.6$\%$  & 2.0$\%$   & 0.5$\%$   & 0.9$\%$    & 0.2$\%$    \\\hline
  \end{tabular}
  \caption{Unweighting efficiencies obtained with the integration in AMEGIC++.}
  \label{Tab:2}
  }
  \end{center}
\end{table}     
\noindent
The phase space integration and the corresponding unweighting efficiencies rely on an 
integrator based on a hierarchical antenna generation (HAAG) \cite{vanHameren:2002tc},
which has been further improved with VEGAS \cite{Lepage:1980dq}.  Results for the unweighting
efficiencies are displayed in Tab.\ \ref{Tab:2}.

\section{From parton to hadron level}

\noindent
For experimental analyses, however, parton level results, discussed in the previous section, 
are of limited interest only.  This is due to the fact that the experimental discussion of 
jets is based on hadrons rather than on partons.  At the moment, the transition from partons 
to hadrons can be described with phenomenological models only, which depend on tunable parameters.  
In order to guarantee the validity of such a tuned parameter set, the partons entering these
models should have comparable distances in phase space.  Ultimately, this is what the parton 
shower (PS), modelling secondary Bremsstrahlung emissions, is responsible for.  

\noindent
When comparing matrix elements (MEs) with the PS, it becomes apparent that they perform best in 
different regimes of particle creation.  While MEs essentially are well-suited to describe hard, 
large-angle emissions, taking interferences into account, the PS covers especially soft and collinear 
emissions, resumming corresponding large logarithms.  It is therefore natural to try to combine 
both descriptions into a unified one, employing the best of both approaches for an improved 
simulation.  The catch in so doing is to avoid double-counting of emissions into the same region of
phase space and to preserve the correct treatment of leading logarithms.

\noindent
An algorithm satisfying these requirements has been presented in \cite{Catani:2001cc} for the case 
of $e^+e^-\to$ hadrons, where its accuracy up to next-to leading logarithmic order has been proven.  
An extension to hadronic collisions has been presented in \cite{Krauss:2002up}.  This algorithm aims 
at a description of all jet emissions correct at tree level plus leading logarithms, with all soft 
and collinear emissions correctly taken care of in the PS.  To achieve this, parton emission is 
separated into two regimes, one for jet production and one for jet evolution, through a 
$k_\perp$-algorithm \cite{Catani:1991hj}.  Jets are then produced according to tree level MEs, 
the corresponding configurations are re-weighted with analytical Sudakov form factors and running 
$\alpha_s$ weights.  In the PS, production of additional hard jets is vetoed.  Altogether, this 
algorithm has been implemented in a process-independent way in \sherpa \cite{Gleisberg:2003xi}, 
allowing for careful validation and cross checks with experimental data or other 
calculations \cite{Krauss:2004bs}.

\begin{figure}
\begin{center}
\begin{tabular}{cc}
\begin{minipage}[ht]{6cm}
  \includegraphics[width=5cm]{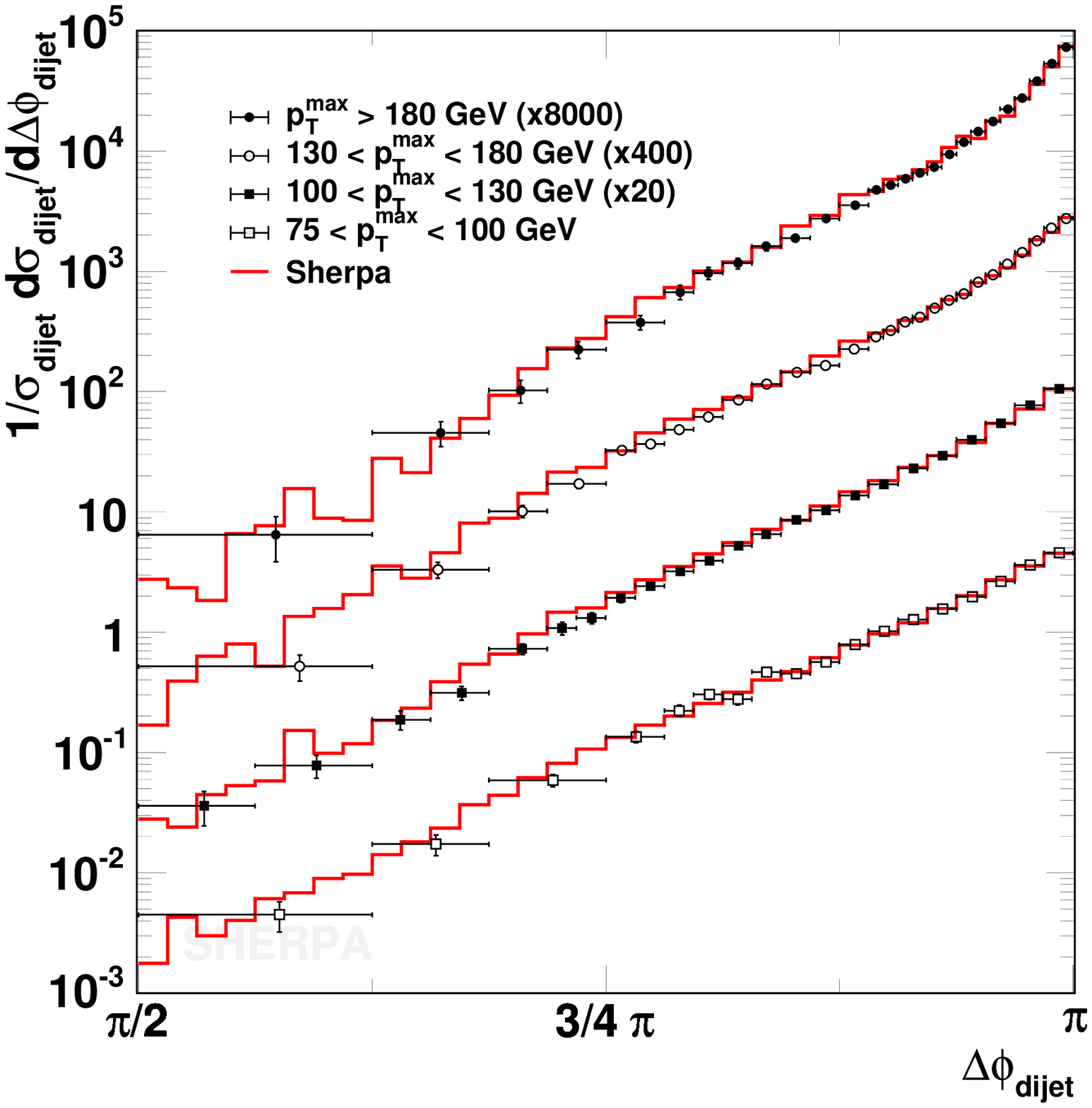}
\end{minipage}
&
\begin{minipage}[ht]{6cm}
  \includegraphics[width=5cm]{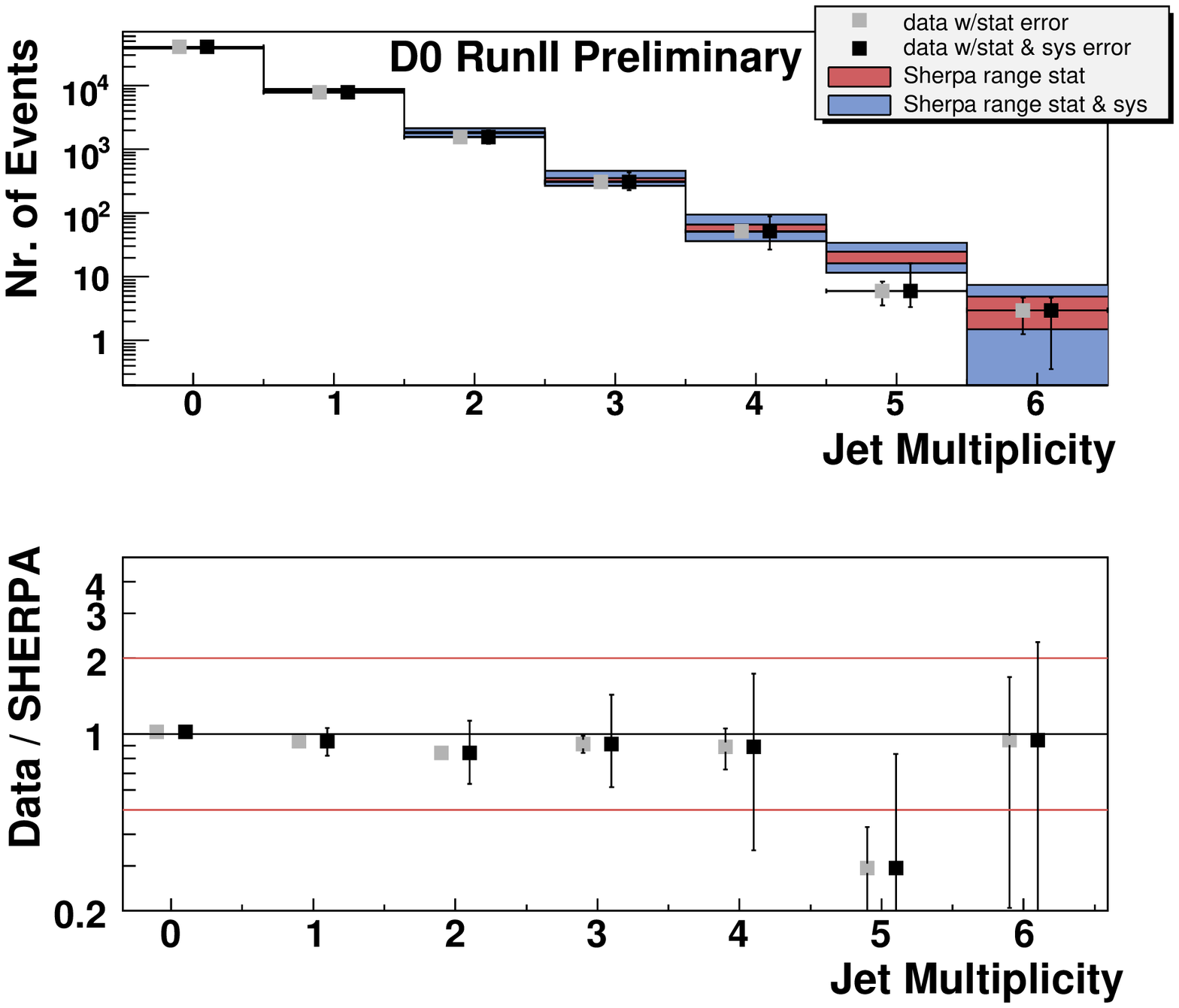}
\end{minipage}\\
\begin{minipage}[ht]{6cm}
  \includegraphics[width=5cm]{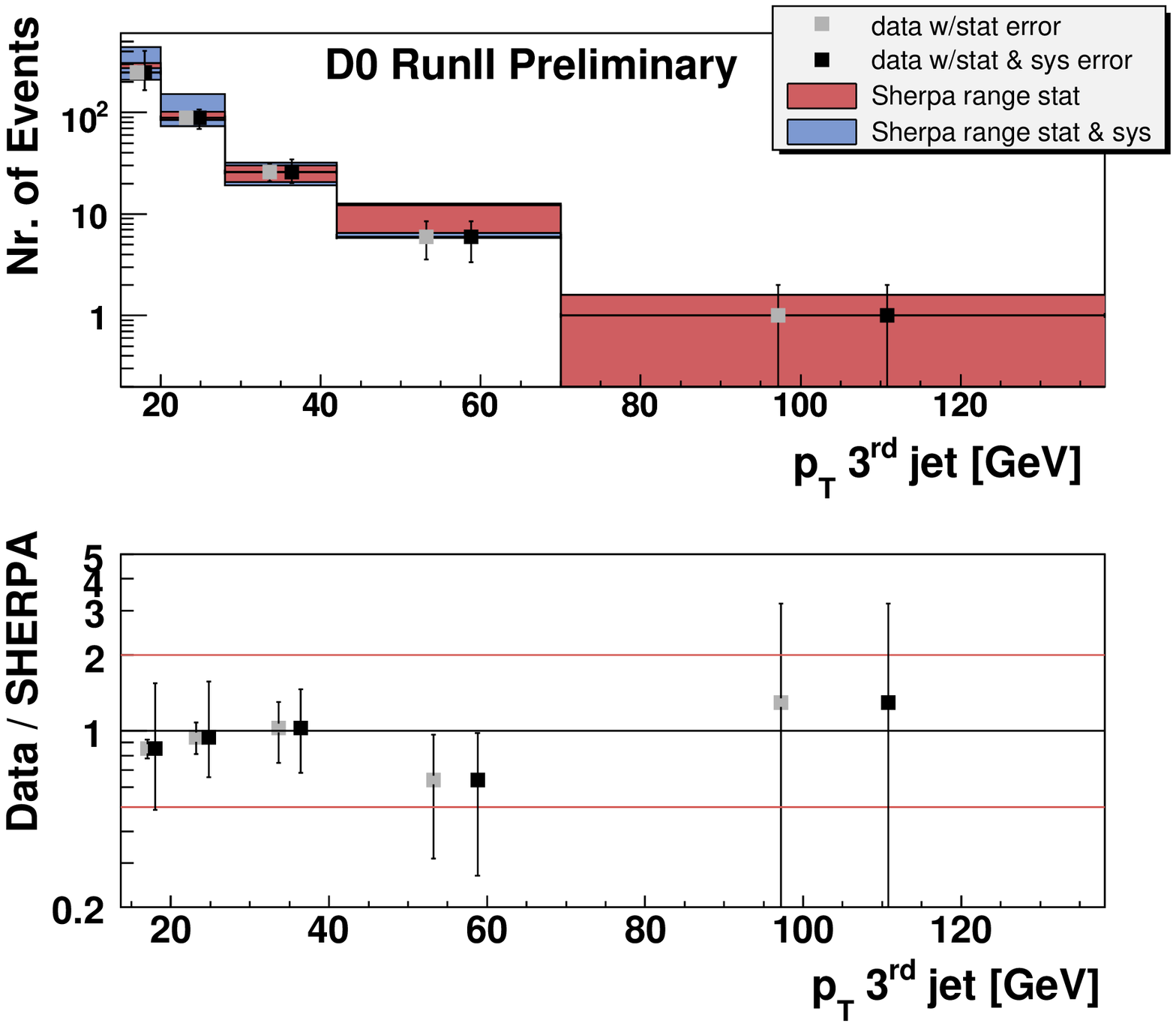}
\end{minipage}
&
\begin{minipage}[ht]{6cm}
  \includegraphics[width=5cm]{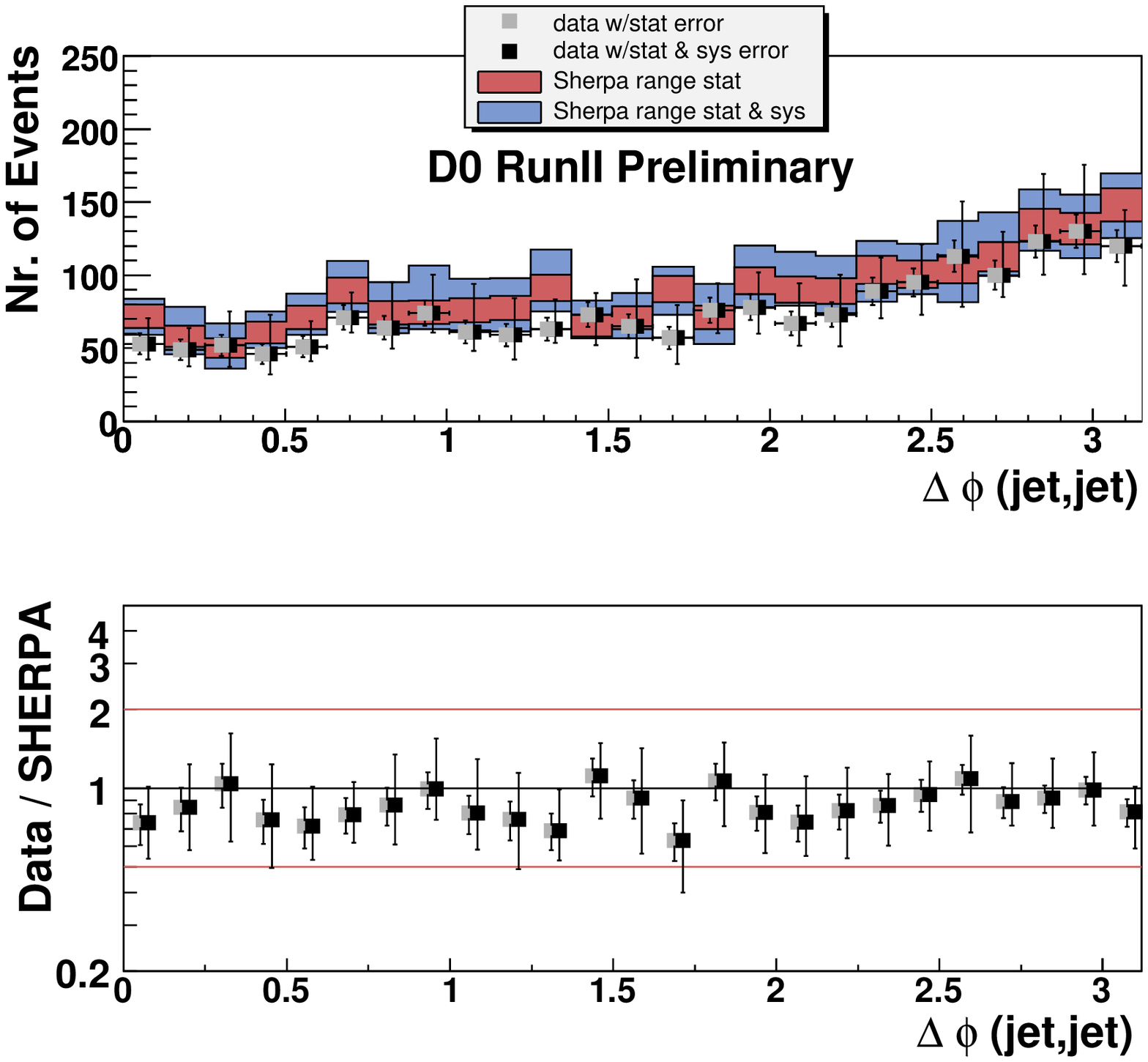}
\end{minipage}
\end{tabular}
  \caption{The azimuthal decorrelation of jets in QCD events (upper left panel), 
           and jet multiplicities (upper right panel), the transverse momentum of the 
           third-hardest jet (lower left panel) and the azimuthal correlation of the 
           two leading jets (lower right panel) in associated $Z$+jet production. 
           All measurements by D\O\ at Tevatron, Run II, and compared with the results 
           from SHERPA.}
  \label{Fig:D0results1}
\end{center}
\end{figure}

\noindent
As a non-trivial check of the quality in describing the QCD radiation pattern through the merging 
approach, consider the azimuthal decorrelation of jets in $p\bar p$ collisions at the Tevatron, Run II, 
as presented in \cite{Abazov:2004hm}.  This observable effectively tests additional radiation, both 
hard and soft, in inclusive QCD dijet production.  The agreement of the results of a \sherpa 
simulation with the experimental results is remarkable, cf.\ the upper left panel of 
Fig.\ \ref{Fig:D0results1}.  A prime example for the predictive power of the merging approach of 
\sherpa is the case of inclusive $Z$ as measured by the D\O\ collaboration at Tevatron, Run II
\cite{henrik}.  There, \sherpa is not only perfectly capable to predict the relative 
multiplicities of associated jets, cf.\ the upper right panel of Fig.\ \ref{Fig:D0results1};
it also yields an improved  description of the jet kinematics. This is illustrated in 
the lower panels of Fig.\ \ref{Fig:D0results1}, where the transverse momentum distribution 
of the third-hardest jet (left) and the azimuthal correlation of the two leading jets 
(right) are displayed.  In so doing, its abilities stretch beyond those of other, more 
traditional event generators, which do not rely on such a merging approach.  

\section{Modelling hadron decays}

\begin{figure}
\begin{center}
\begin{tabular}{cc}
\begin{minipage}[ht]{6cm}
  \includegraphics[width=5cm]{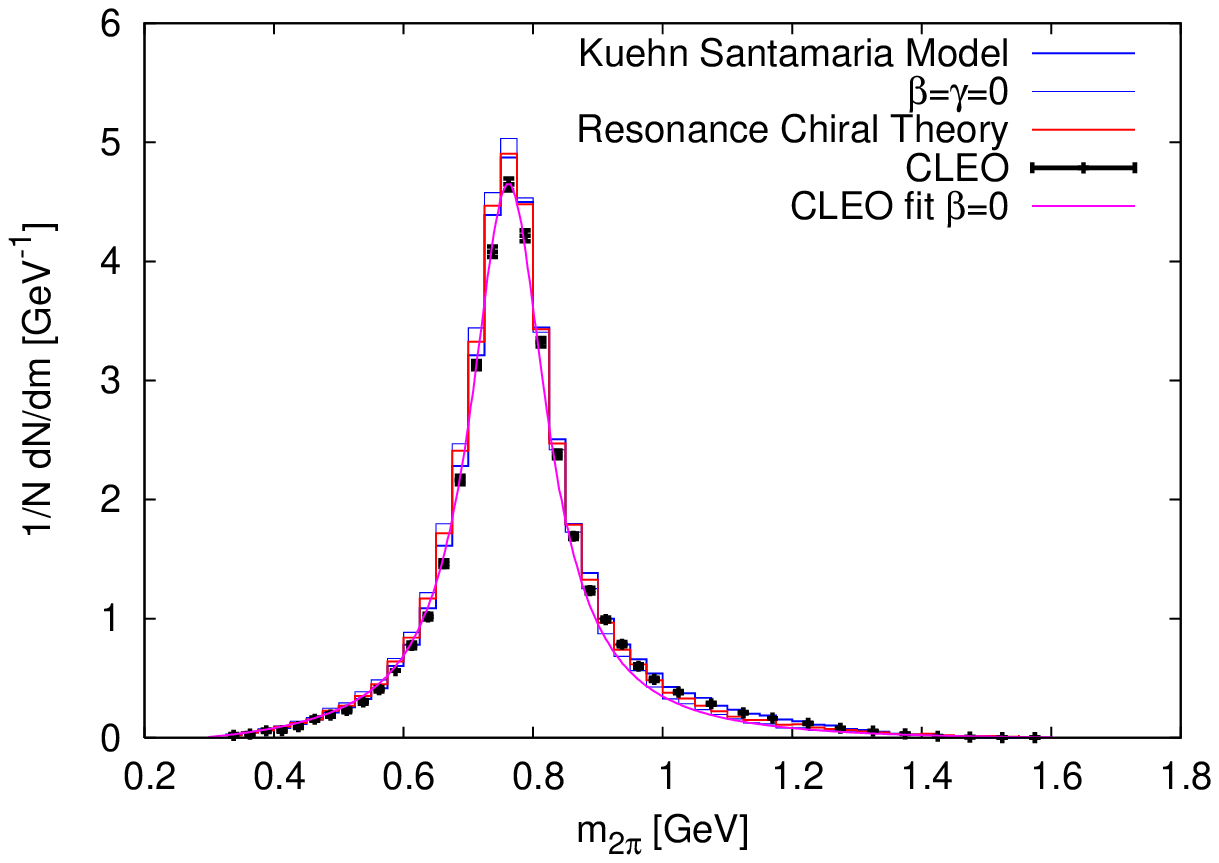}
\end{minipage}
&
\begin{minipage}[ht]{8cm}
  \includegraphics[width=8cm]{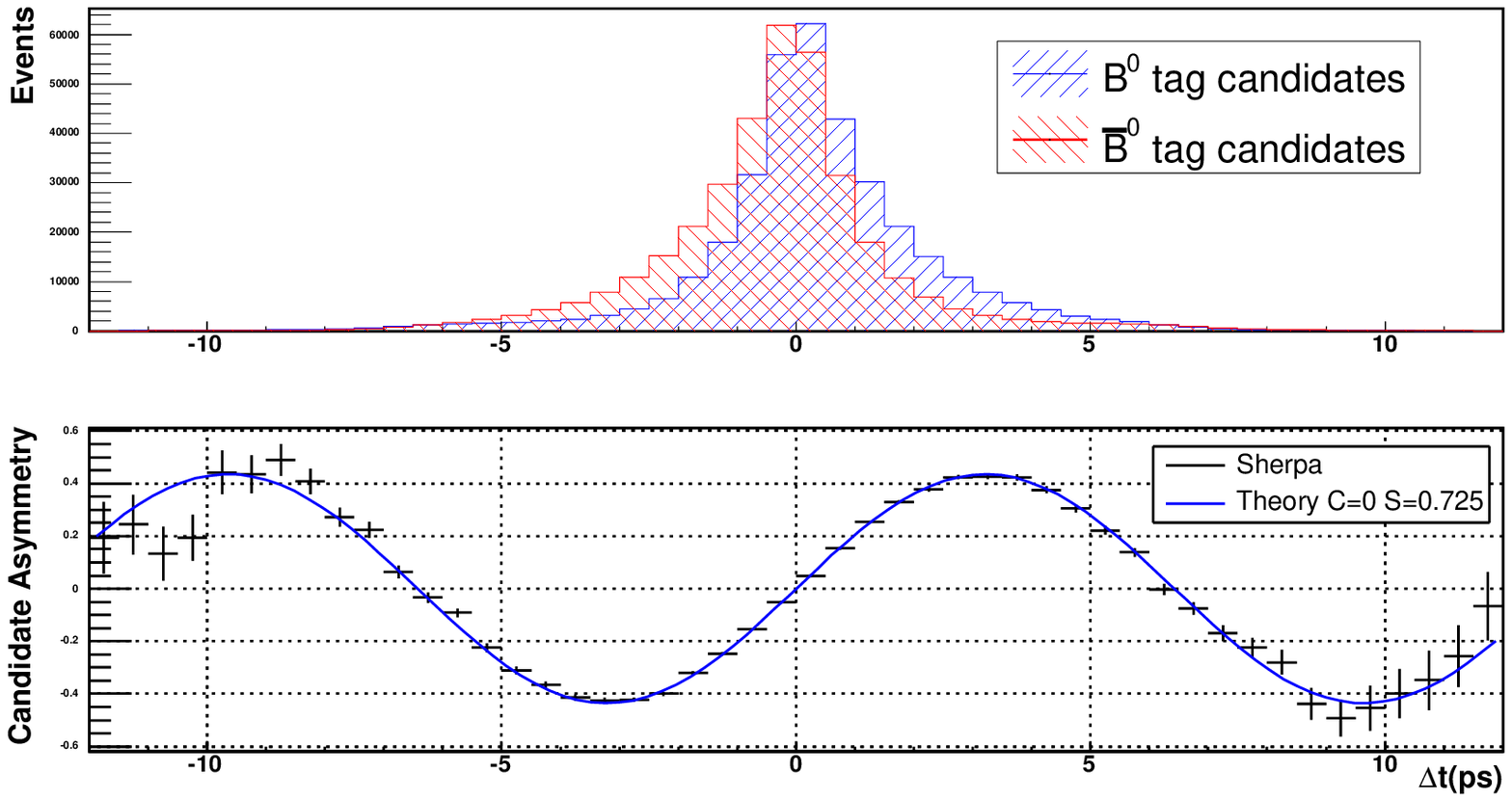}
\end{minipage}
\end{tabular}
  \caption{The impact of different form factor models on the decay 
           $\tau\to\pi\pi\nu_\tau$ (left panel) and the simulation of
           $B\bar B$ mixing on the asymmetry of $J/\Psi K_S$ final states
           in $B$ decays (right panel); results from SHERPA.}
  \label{Fig:Hadron}
\end{center} 
\end{figure}

\noindent
Another improvement of modern event generators when compared to traditional ones rests in the 
description of hadron decays and decay chains.  Apart from the inclusion of spin correlations 
\cite{Richardson:2001df}, modelling the effect of interferences in decay chains, apparent refinement 
of the simulation can be achieved by using better form factor models in decays, leading to 
non-flat phase space distributions and by an upgraded description of mixing effects, like, 
e.g.\ $B\bar B$ mixing.  Some of these refinements are exemplified in Fig.\ \ref{Fig:Hadron}.  
There, in the left panel, the effect of different form factor models \cite{formfactors} on 
$m_{\pi\pi}$ in decays $\tau\to\pi\pi\nu_\tau$ are compared with experimental data from 
\cite{Anderson:1999ui}, whereas in the right panel the asymmetry of $J/\Psi K_S$ final states
in $B$ decays is displayed.

\section{Conclusions}

\noindent
In this contribution, the need for new simulation tools in preparation for a successful LHC era 
has been motivated.  These new tools become mandatory due to the abundance of backgrounds, shadowing 
potentially interesting signals.  An apparent feature of modern event generators, improving 
traditional ones rests in the systematic inclusion of higher-order QCD corrections through merging 
or matching algorithms.  One of them has been shortly discussed, and results obtained with it have 
been presented.  In order to realise tree-level merging algorithms, multi-leg tree level parton 
level event generators are an important ingredient, and some recent developments concerning the 
efficient calculation of corresponding cross sections have been shown.  Finally, another rectification 
included in modern tools, consists in a better understanding and modelling of hadron (especially $B$ 
and $D$) and $\tau$ decays and in the simulation of non-trivial quantum interference effects.

\section*{References}

\end{document}